\documentclass[twocolumn,prl,superscriptaddress]{revtex4}
\usepackage{graphicx}% Include figure files
\usepackage{dcolumn}% Align table columns on decimal point
\usepackage{bm}% bold math
\usepackage{color}
\usepackage{ulem}

\begin{document}

% Use the \preprint command to place your local institutional report
% number in the upper righthand corner of the title page in preprint mode.
% Multiple \preprint commands are allowed.
% Use the 'preprintnumbers' class option to override journal defaults
% to display numbers if necessary
%\preprint{}

\title{Quantum critical dynamics of a $\bf S = 1/2$ antiferromagnetic
Heisenberg chain studied by $^{13}$C NMR spectroscopy}

\author{H. K\"{u}hne}
\author{H.-H. Klauss}
\affiliation{Institut f\"{u}r Festk\"{o}rperphysik, TU Dresden,
01069 Dresden, Germany
}%%%
\affiliation{Institut f\"{u}r Physik der Kondensierten Materie, TU
Braunschweig, 38106 Braunschweig, Germany
}%%%
\author{S.Grossjohann}
\author{W. Brenig}
\affiliation{Institut f\"{u}r Theoretische Physik, TU Braunschweig,
38106 Braunschweig, Germany
}%%%
\author{F.J. Litterst}
\affiliation{Institut f\"{u}r Physik der Kondensierten Materie, TU
Braunschweig, 38106 Braunschweig, Germany
}%%%
\author{A.P. Reyes}
\author{P.L. Kuhns}
\affiliation{National High Magnetic Field Laboratory, Tallahassee,
Florida 32310, USA
}%%%
\author{M.M. Turnbull}
\author{C.P. Landee}
\affiliation{\mbox{Carlson School of Chemistry and Department of
Physics, Clark University, Worcester, Massachusetts 01610, USA}
}%%%

\date{\today}

\begin{abstract}
We present a $^{13}$C-NMR study of the magnetic field driven
transition to complete polarization of the S=1/2 antiferromagnetic
Heisenberg chain system copper pyrazine dinitrate
Cu(C$_4$H$_4$N$_2$)(NO$_3$)$_2$ (CuPzN). The static local
magnetization as well as the low-frequency spin dynamics, probed via
the nuclear spin-lattice relaxation rate $T_1^{-1}$, were explored
from the low to the high field limit and at temperatures from the
quantum regime ($k_B T\ll J$) up to the classical regime ($k_B T\gg
J$). The experimental data show very good agreement with quantum
Monte Carlo calculations over the complete range of parameters
investigated. Close to the critical field, as derived from static
experiments, a pronounced maximum in $T_1^{-1}$ is found which we
interpret as the finite-temperature manifestation of a diverging
density of zero-energy magnetic excitations at the field-driven
quantum critical point.
\end{abstract}

% insert suggested PACS numbers in braces on next line
\pacs{75.10.Jm, 75.50.Ee, 75.30.Gw, 75.50.Xx}
% insert suggested keywords - APS authors don't need to do this
%\keywords{}

%\maketitle must follow title, authors, abstract, \pacs, and \keywords
\maketitle

\section{I. Introduction}

Quantum critical points (QCPs), i.e., zero-temperature phase
transitions as a function of some control parameter, are likely to
be at the core of unconventional finite-temperature behavior of many
novel materials \cite{Chakravarty1989a,Chubukov1994a}. Following the
pioneering analysis of spin chains
\cite{Affleck1990_91,Sorensen1993} and spin ladders
\cite{Giamarchi1999} in external magnetic fields, Bose-Einstein
condensation of hard-core bosons has been related to some phase
transitions in quantum magnets which stem from the level crossing of
elementary triplet excitations with the ground state at some
critical external magnetic field $B=B_c$.

Field-induced QCPs have been under intense scrutiny for three and
quasi-two-dimensional spin $S=1/2$ dimer systems, i.e., TlCuCl$_3$
\cite{Nikuni2000,Oosawa2001,Ruegg2003,Misguich2004} and
BaCuSi$_2$O$_6$ \cite{Sebastian2006}, for $S=1/2$ ladder materials
Cu$_2$(C$_5$H$_12$N$_2$)$_2$Cl$_4$ \cite{Chaboussant1997} and
(C$_5$H$_{12}$N)$_2$CuBr$_4$ \cite{Watson2001,Lorenz2008}, for the
$S=1$ Haldane chain Ni(C$_5$H$_{14}$N$_2$)$_2$N$_3$(PF$_6$)
\cite{Honda1998}, for the coupled chain compound
NiCl$_2$-4SC(NH$_2$)$_2$ (DTN) \cite{Zapf2006,Zvyagin2007} with
$S=1$, as well as for the effective $S=1$ system
(CH$_3$)$_2$CHNH$_3$CuCl$_3$ \cite{Manaka1998,Garlea2007}. All of
the latter materials feature a gapful zero-field state with the
lowest triplet branch condensing as the field is {\it increased}.
However, a similar scenario can be realized in the antiferromagnetic
$S=1/2$ Heisenberg chain (AFHC) upon {\it decreasing} the field
through the critical value for complete polarization $B_c$. The
Hamiltonian of the AFHC in an external field reads
\begin{equation}\label{eqn1}
H = \sum_{i} (J \mathbf{S}_i \cdot \mathbf{S}_{i+1} - g \mu_B
\mathbf{B} \cdot \mathbf{S}_i),
\end{equation}
where ${\bf S}_i$ are spin operators and $J$ is the exchange energy.
For $B>B_c=2J/(g\mu_B)$ the lowest elementary excitation is a single
Ising triplet which crosses the ground state at $B=B_c$, where the
system switches from complete polarization into a Luttinger liquid
of deconfined spinons. As for other one-dimensional (1D) systems
investigated, i.e., Haldane chains and spin ladders, true gauge
symmetry breaking for the triplet bosons will not occur at $B_c$,
however power-law correlations will develop, which are manifested in
the spin-correlation functions \cite{Affleck1990_91,Sorensen1993}.

Previous studies of field-driven criticality in quantum magnets have
been focused on thermodynamic properties. The dynamics remain a
rather open issue. Therefore, the purpose of this letter is to shed
light on the field-induced spin-dynamics of the AFHC. We report
results of a nuclear magnetic resonance (NMR) study of the low
frequency spin response for a wide range of parameters from low
fields ($B \ll B_c$) to the high field limit ($B \gg B_c$), as well
as from the quantum regime ($k_BT\ll J$) to the classical regime
($k_BT\gg J$). On the one hand, the dynamics are probed by the
nuclear spin-lattice relaxation rate $T_1^{-1}$ measured in the
metalorganic AFHC CuPzN. On the other hand, the experimental data
are compared to quantum Monte Carlo (QMC) calculations. Additionally
we also investigated the static magnetic properties by comparison of
the NMR frequency shift $\delta$ with the magnetization calculated
by QMC. We find very good agreement between experiment and theory in
all cases.

\section{II. Experimental}

\begin{figure}
\begin{center}
\includegraphics[width=1.0\columnwidth]{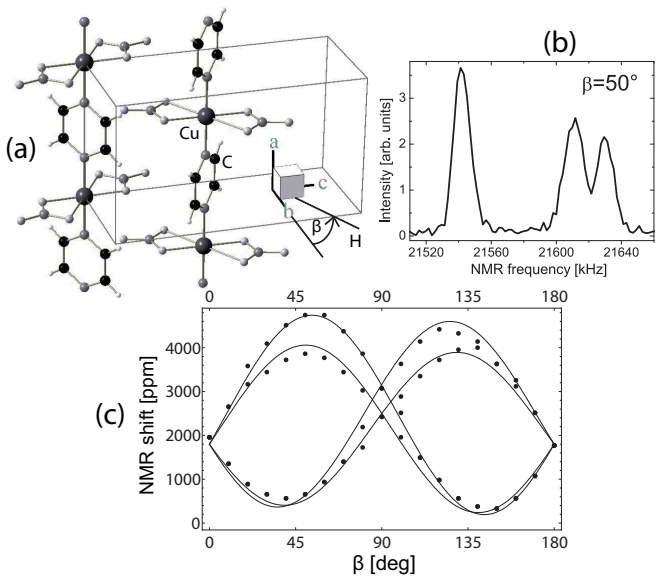}
\end{center}
\caption[1]{(a) Crystal structure of CuPzN. (b) $^{13}$C NMR
spectrum for $B_0= 2$ T, $T=7$K, and $\beta = 50^\circ $ in the
$b$-$c$ plane. (c) NMR center shift $\delta$ of spectral lines
versus $\beta$ compared with a simulation of the local magnetic
field ~\cite{simshift}.} \label{fig1}
\end{figure}

The compound CuPzN, i.e., copper pyrazine dinitrate
Cu\-(C$_4$\-H$_4$\-N$_2$)\-(N\-O$_3$)$_2$, is one of the best
realizations of the AFHC. Compared to oxide-based AFHC systems
\cite{Ishida1994,Motoyama1996,Ami1995,Takigawa1996,Takigawa1997} it
has a small exchange-coupling constant $J/k_B=10.7$ K which allows
experimental access to the parameter range of the saturation field
$B_c\sim 14.9$ T and above ~\cite{losee}. The unit cell is
orthorhombic, at room temperature the lattice constants are
$a=6.712$ {\AA}, $b=5.142$ {\AA}, and $c=11.732$
{\AA}~\cite{santoro}, see Fig. \ref{fig1}(a). The 1D chains are
equally spaced, their axis being parallel to $a$. The Cu(II) ions on
a chain are separated by pyrazine rings which mediate the
antiferromagnetic coupling between the copper moments mainly via
superexchange. Recently, three-dimensional ordering was observed at
107 mK ~\cite{lancaster}. The ratio $ \mid
J_{interchain}/J_{intrachain} \mid = 4.4 \times 10^{-3} $ indicates
the highly one-dimensional character of this system. CuPzN has been
characterized by inelastic neutron scattering, muon-spin relaxation,
magnetothermal transport, specific heat, and magnetization
measurements ~\cite{hammar,Stone03,lancaster,solo}. All of these
studies are consistent with a description
of CuPzN in terms of the AFHC.\\
Single crystals of CuPzN have been grown as described previously
~\cite{hammar}. The crystal used for the measurements presented in
this letter has the dimensions $2.6 \times 4.2 \times 0.8$ mm$^3$
and a mass of $m$ = 9.09 mg. The $^{13}$C nucleus in the pyrazine
ring was used as the $I$=1/2 NMR probe since the copper nuclei yield
an experimentally very unfavorable spin-spin relaxation time
$T_2<10$ $\mu s$. The measurements at $^{13}$C were done for several
fields between 2 and 28 T and temperatures between 1.5 and
50 K.\\
The low-field data were recorded in an 8 T superconducting magnet
with a modified Bruker CXP200 spectrometer, applying a standard
inversion-recovery spin-echo pulse sequence. The measurements at
higher fields were done at the NHMFL, Tallahassee, in a 17 T
superconducting magnet and a 31 T resistive magnet with a home-built
spectrometer, using a Carr-Purcell-Meiboom-Gill (CPMG)
pulse-sequence for $B>20$ T.

\section{III. Nuclear Magnetic Resonance}

The nuclear spin-lattice relaxation rate $T_1^{-1}$ measures the
spin fluctuations at the nuclear Larmor frequency $\omega_n$
~\cite{mor},
\begin{eqnarray}
\frac{1}{T_1}= \frac{\gamma_n^2}{2} \sum_{q} \sum_{\beta = x,y,z} [A
_{x \beta}^2 (q) + A _{y \beta}^2 (q)]\hspace{1.5cm} \nonumber
\\ \times \int_{- \infty} ^{\infty} < S_{\beta}
(q,t)S_{\beta} (-q,0)> e^{-i \omega_n t} dt
\label{eqn2} \\
=\frac{\gamma_n^2}{2} \sum_{q}[F_{\perp}(q) S_{\perp}(q, \omega_n) +
F_{z}(q) S_{z}(q, \omega_n)]. \label{eqn3}
\end{eqnarray}
Here, $F_{\perp}(q)$ and $F_{z}(q)$ are the geometrical form factors
and $S_{\perp}(q, \omega)$ and $S_{z}(q, \omega)$ are the dynamical
structure factors of the electronic spin system. $A_{\alpha \beta}$
with $\alpha, \beta=x,y,z$ are the components of the hyperfine
coupling tensor $\underline{\mathbf{A}}(q)$. In CuPzN the $^{13}$C
nuclei are coupled to the magnetic moments of the Cu(II) electrons
via isotropic hyperfine coupling $A_{iso}(q)$, mediating only
transverse spin fluctuations, and anisotropic dipolar coupling
$\underline{\mathbf{A}}_{dip}(q)$, mediating transverse and
longitudinal spin fluctuations ~\cite{azevedo}.
\begin{figure}
\begin{center}
\includegraphics[width=1.0\columnwidth]{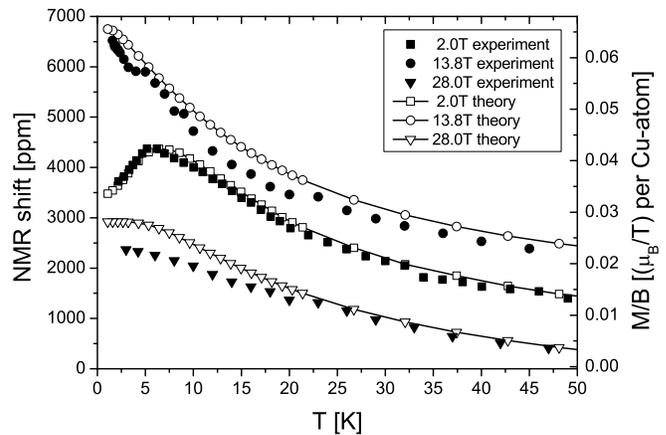}
\end{center}
\caption[1]{Comparison of temperature-dependent NMR shift $\delta$
with magnetization data calculated by QMC. For clarity, an offset
has been added to the data at 13.8 T (+1000 ppm to NMR shift and
0.0099 ($\mu_B$ / T) to calculated magnetization) and 28 T (-1000
ppm and -0.0099 \linebreak ( $\mu_B$ / T), respectively). The QMC
errors are within symbol size. All solid lines are a guide to the
eye. } \label{fig2}
\end{figure}
We want to compare the experimentally and theoretically determined
$T_1^{-1}$ rates for transverse fluctuations. Therefore, the dipolar
contribution to $\underline{\mathbf{A}}(q)$ has to be minimized.
This minimum is found for the orientation $\beta= 50^\circ$ via a
study of the angular dependence of the NMR shift $\delta =(
\omega_{n}-\gamma B_0 )  / \gamma B_0$ , see Figs. \ref{fig1}(b) and
\ref{fig1}(c) ~\cite{wolter, simshift}.

\subsection{A. NMR frequency shift}

For a fixed orientation of the external field the NMR shift $\delta$
is related to the magnetization $M(T)$ via
\begin{eqnarray}
\delta(T) & = & A(q=0) \cdot \frac{M(T)}{B}.
\end{eqnarray}
The shift $\delta$ is compared with the calculated magnetization of
a S=1/2 AFHC  in Fig. \ref{fig2}, scaling the latter with the same
factor $A(q=0)=0.101$ T $/ \mu_B $  for all fields. $A(q=0)$ was
determined by a least-squares fit of the 2 T data sets. For 2 T we
find excellent agreement between experiment and theory, both showing
a broad maximum around 6.5 K, reflecting the onset of
antiferromagnetic correlations. At 13.8 T, slightly below the
saturation field, both data sets show monotonous increase toward
saturation with decreasing temperature. The kink near 5 K in the
experimental data is due to the proximity of the boiling point of
liquid Helium. At 28 T, experiment and theory deviate below 20 K. An
rf heating of the sample can be excluded since the conditions of the
CPMG pulse-sequence were carefully adjusted.

\subsection{B. Transverse dynamic structure factor}

Before turning to the $T_1^{-1}$ data, we present our method of
calculation for the field and temperature-dependent transverse
dynamic structure factor $S_{\perp}(q,\omega_n)$. Switching to
imaginary time $\tau$ the latter reads  $S_{\perp}(q,
\tau)=\frac{1}{\pi}\int_{0}^{\infty}d\omega S_{\perp}(q,
\omega)K(\omega, \tau)$, with a kernel $K(\omega,
\tau)=e^{-\tau\omega} + e^{-(\beta-\tau)\omega}$ and $\beta=1/T$. In
real-space $S_{\perp}(q,\tau)$ can be calculated efficiently, using
QMC. Following Ref. \cite{Sandvik1992}
\begin{eqnarray}\label{eqn5}
S_{i, j}\left( \tau\right)
=\left\langle \sum_{p,m=0}^{n}
\frac{\tau^m(\beta-\tau)^{n-m}n!}{\beta^n(n+1)(n-m)!m!} \times
\right . \phantom{a a a}
&&\nonumber\\
\left . S^{+}_{i}(p)S^{-}_{j}(p+m) \right\rangle_W ~, &&
\end{eqnarray}
where $S_{\perp}(q, \tau)=\sum_a e^{iq a}S_{a,0}(\tau)/N$ and $a,0$
label lattice sites in a chain of length $N$.
$\langle\ldots\rangle_W$ refers to the Metropolis weight of an
operator string of length $n$ generated by the stochastic series
expansion of the partition function
\cite{Sandvik1999a,Syljuaasen2002}, and $p,m$ are positions in this
string.

Analytic continuation from imaginary times, i.e., $S_{\perp}(q,
\tau)$, to real frequencies, i.e., $S_{\perp}(q, \omega)$, is
performed by the maximum entropy method (MaxEnt), minimizing the
functional $Q=\chi^2/2 - \alpha\sigma$ \cite{bryan1,Jarrell1996}.
Here $\chi$ refers to the covariance of the QMC data to the MaxEnt
trial-spectrum $S_{\alpha\perp}(q, \omega)$. Overfitting is
prevented by the entropy $\sigma = \sum_{\omega}S_{\alpha\perp}(q,
\omega)\ln[S_{\alpha\perp}(q, \omega)/m(\omega)]$. We have used a
flat default model $m(\omega)$, matching the zeroth moment of the
trial spectrum. The optimal spectrum follows from the weighted
average
\begin{equation}
S_{\perp}(q, \omega) = \int_{\alpha}d\alpha P[\alpha | S(q,
\tau)]S_{\alpha\perp}(q, \omega) ~,
\end{equation}
with the probability distribution $P[\alpha | S(q, \tau)]$ adopted
from Ref. \cite{bryan1}.

\begin{figure}
\begin{center}
\includegraphics[width=1.0\columnwidth]{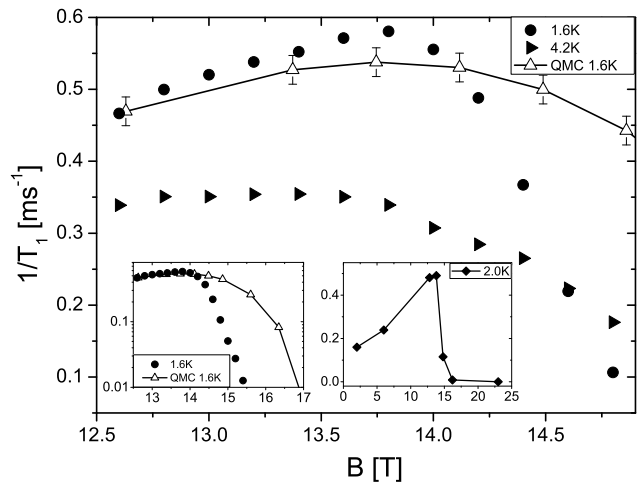}
\end{center}
\vskip -.5cm \caption[1]{Field dependence of the nuclear
spin-lattice relaxation rate of $^{13}$C in the critical regime.
Left inset: The log-scale plot demonstrates the linear opening of
the spin gap with field. Right inset: The full-scale plot highlights
the maximum of $T_1^{-1}(B)$ near the $T$=0 K critical field. All
solid lines are a guide to the eye.} \label{fig3} \vskip -.5cm
\end{figure}

\subsection{C. $\mathbf{1/T_1}$: experiment versus theory}

Turning to the form factors in Eq. (\ref{eqn3}), Fig. \ref{fig1}(a)
shows that the NMR site, i.e., the carbon nucleus, is located
asymmetrically between two Cu(II) ions. Therefore on-site and
next-nearest-neighbor correlations are included by using an
effective real space form factor $F(r) =  [F_0 \delta(r) + F_1
\delta(r-a)]$, where $a$ is the lattice constant and $F_{0,1}$
parameterize the hyperfine coupling between the nucleus and its
nearest copper moments. This leads to a transverse relaxation rate
of
\begin{equation}
\frac{1}{T_1} = F_0^2\left[(1+R^2)S_\perp(0,\omega) +
2RS_\perp(1,\omega)\right]|_{\omega\rightarrow 0},
\end{equation}
where $S_\perp(r=0(1),\omega)$ are the real-space transverse-spin-
correlation functions at a distance $r=0(1)$.

In Fig. \ref{fig3} we compare the observed NMR rate with the QMC
results versus magnetic field in the quantum regime $k_BT\ll J$,
with $T=1.6$ K. The QMC data is shown for $R=0$ and a single overall
scaling factor $F_0$ assigned at 2 T and high temperatures. The
similarity between experiment and theory is remarkable. For both we
find a pronounced maximum of $T_1^{-1}(B)$ at $B=13.8$ T shifting to
lower fields with increasing temperature. To interpret these
results, we note that in the fully polarized state for $B>B_c$,
single magnons are exact eigenstates of Eq. (\ref{eqn1}) with a
dispersion of
\begin{eqnarray}
E_>(k) = J \cos (k) + g \mu_B B~,
\end{eqnarray}
$E_>(k)$ displays a field-driven excitation gap of $g \mu_B B-J$
leading to an exponential decrease in $T_1^{-1}(B)$ at fixed $T$ and
for $B>B_c$. This can be seen for both, NMR and QMC, on the
$\log$-scale left inset in Fig. \ref{fig3}. The rates calculated by
QMC display a broader maximum than the measured data, but drop with
the same slope for fields above $16$ T. We emphasize that this
deviation between NMR data and QMC is confined to low temperatures
$T\lesssim 6$ K and to a limited range of fields $15$ T $\lesssim B
\lesssim 17$ T which can be seen from the log-scale of the left
inset. At $B=B_c$ the dispersion touches the zero at $k=\pi/2$ with
a {\it quadratic} momentum dependence yielding a van-Hove type of
critical DOS. This leads to the maximum in $T_1^{-1}$, tending to
diverge as $T\rightarrow 0$. For both, NMR experiment and QMC, the
maximum in Fig. \ref{fig3} occurs at $\tilde{B}_c\approx 13.8 $ T,
which is slightly less than the saturation field of $B_c=14.9$ T for
the magnetization. Most likely this downshift is a finite-
temperature effect of excitations populating the gap. In the
Luttinger liquid for $B<B_c$ the low-energy spinon excitations have
a field-dependent {\it linear} dispersion, yielding a finite, yet
reduced NMR rate.

\begin{figure}
\begin{center}
\includegraphics[width=1.0\columnwidth]{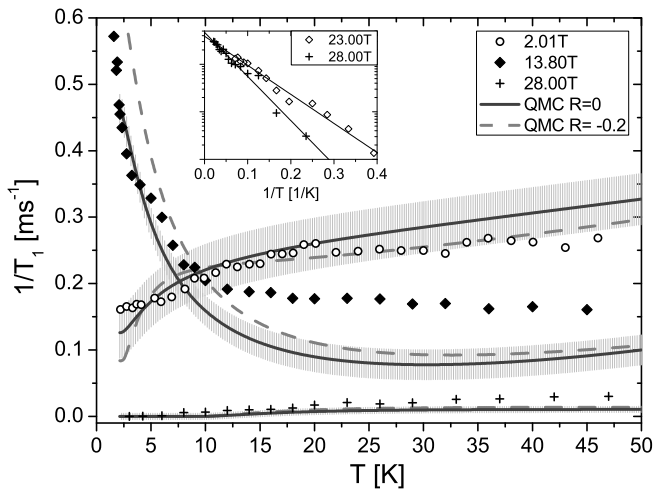}
\end{center}
\vskip -.5cm \caption[1]{Temperature dependence of the nuclear
spin-lattice relaxation rate of $^{13}$C for different external
fields. The solid QMC data lines are each polynomial fits to 50
analytic continuations of a 128 site system and the error tube was
chosen to contain all data points within a range of 2 $\sigma$. The
log-scale inset shows the exponential decrease in $1/T_1$ with $1/T$
above $B_c$.} \label{fig4} \vskip -.5cm
\end{figure}

In Fig. \ref{fig4} we compare $T_1^{-1}$ rates observed
experimentally with QMC results versus temperature for three fields,
i.e., above, at, and below $\tilde{B}_c$. As for the field
dependence, the agreement between theory and experiment is very
good. Inclusion of next-nearest neighbor hyperfine couplings, i.e.,
R$=-0.2$, can slightly improve this agreement at high temperatures
but decreases the agreement at low temperatures. The main result of
this figure is the diverging NMR rate at $\tilde{B}_c$ which is very
suggestive of critical scattering as $T\rightarrow 0$. As $T$
increases, the van-Hove singularity in the DOS at $B_c$ is smeared
leading to the decrease in $T_1^{-1}$. For $B=28$ T $>B_c$, the rate
has dropped by $\sim$ 3 orders of magnitude due to the spin-gap and
increases with temperature following $1/T_1 \propto exp(-\Delta /
k_B T)$. The corresponding fits at 23 and 28T, shown in the inset of
Fig. 4, give $ \Delta_{23T} = 9.6T \pm 0.6T $ and $ \Delta_{28T} =
14.3T \pm 0.9T $, confirming that the gap increases with $g \mu_B
B-J$. Finally, for $B<B_c$ we observe only a weak overall $T$
dependence. In the classical regime $k_BT\gg J$ the rate is
decreasing with increasing fields. This is indicative of an
excitation spectrum dominated by spin-diffusion modes from $q=0$.

\section{IV. Conclusion}

To summarize, by a complementary analysis of experiment and theory
for the low-frequency spin spectrum of the AFHC CuPzN, as probed by
the NMR $T_1^{-1}$ rate as well as by the Knight shift, we have
provided clear evidence for critical dynamics close to a
field-induced QCP. Both, experiment and QMC calculations are in good
agreement and show a pronounced maximum in $T_1^{-1}$ in the
vicinity of the saturation field, which tends to diverge as
$T\rightarrow 0$. Moreover, good agreement between theory and
experiment is also found for the magnetization versus temperature
and field, except for a low-$T$ deviation at $28$ T, yet to be
explored. Our findings may be of interest in the context of other
field-induced QCPs as ,e.g., in TlCuCl$_3$
\cite{Nikuni2000,Oosawa2001,Ruegg2003,Misguich2004} or
BaCuSi$_2$O$_6$ \cite{Sebastian2006}.\\

\section{Acknowledgments}

Part of this work was performed at the National High Magnetic Field
Laboratory, supported by NSF Cooperative under Agreement No.
DMR-0084173, by the State of Florida, by the DOE and the DFG under
Grants No. KL1086/6-2 and No. KL1086/8-1. One of us (W.B.)
acknowledges partial support by the DFG through Grant No. BR
1084/4-1 and the hospitality of the KITP, where this research was
supported in part by the NSF under Grant No. PHY05-51164. 


\begin{thebibliography}{99}
\bibitem{simshift}
The $q=0$ hyperfine coupling tensor $\underline{\mathbf{A}}(0)$ can
be disentangled into $\underline{\mathbf{A}}_{dip}(0)$ and
$A_{iso}(0)$ from the angular dependence of the NMR-frequency shift
$\delta = \frac{1}{B^2} [  \mathbf{B} \cdot ((
\underline{\mathbf{A}}_{dip}(0) + A_{iso}(0)) \cdot
\underline{\mathbf{\chi_{el}}} + \underline{\mathbf{\sigma}}) \cdot
\underline{\mathbf{B}} ]$, where $\underline{\mathbf{\chi_{el}}}$ is
the tensor of the electronic susceptibility and
$\underline{\mathbf{\sigma}}$ is the diamagnetic tensor. A
comparison of the measured and simulated $\delta$ [Fig.
\ref{fig1}(c)] yields a minimum of $\delta_{dip}(0)$ for the high
frequency peak close to an angle of $\beta= 50^\circ$. Since
$\underline{\mathbf{A}}_{dip}(q)$ scales with
$\underline{\mathbf{A}}_{dip}(0)$, this orientation with
$\delta_{iso}(0)/ \delta_{dip}(0) \approx 4 $ was used for all
subsequent measurements shown in Figs. 2-4. Note that the simulation
also yields the splitting of the high-frequency peak due to a
misalignment between the rotation axis and the $a$b axis of the
crystal by $\sim$ $5^\circ$.
\end{thebibliography}
\end{document}